\input epsf
\epsfverbosetrue
\documentclass[a4paper,12pt,twoside]{article}
 
\usepackage{graphics}
\include{definitions}
\begin{document}
\leftmargin -2cm

\begin{large}
{\bf Searching for obscured supernovae in nearby starburst galaxies} 
\end{large} 
\vspace{-0.5cm} 
\begin{center}
Seppo Mattila$\footnote{email: s.mattila@ic.ac.uk}$ (Imperial College), Robert 
Greimel (ING), Peter Meikle (Imperial College), Nicholas A. Walton (IoA), 
Stuart Ryder (AAO), Robert D. Joseph (IfA) 
\end{center}
\vspace{-0.1cm} 
We are currently carrying out a $K_{\rm s}$ band survey for core-collapse 
supernovae (CCSNe) in the nuclear 
(central kpc) regions of nearby starburst galaxies with the INGRID near-IR 
camera at the WHT. In this article we concentrate in describing mainly the 
observations and the real time processing of the SN search data, which
makes
use of the ING's integrated data flow system. 

Very little is currently known about the behaviour of SNe in a starburst 
environment.  The enhanced metallicity in starburst regions is expected to 
result in large mass-loss rates (Vink et al. 2001) for the SN progenitor 
stars.  In addition, many of these events are expected to occur within 
molecular clouds (Chevalier $\&$ Fransson 2001), adding further to the 
density of material surrounding the SN. 
Therefore, nuclear SNe can be expected to explode 
within a dense circumstellar medium. In general, a dense (but non-nuclear) 
circumstellar environment has been observed to produce bright CCSNe with slow 
near-IR light decline rates (e.g. SN 1998S, Fassia et al. 2000). However, the 
behaviour of SNe in nuclear starburst regions may be much more extreme 
(Terlevich et al. 1992). The high-${\it z}$ CCSN surveys 
with the VLT, {\it HST} and {\it NGST} (e.g., 
Dahlen \& Fransson 1999; Sullivan et al. 
2000) will use SNe to probe the cosmic star formation rate.  For this, it is 
important to determine (i) a better estimate of the {\it complete} local CCSN rate 
(cf. Sullivan et al. 2000), (ii) a proper understanding of the behaviour of SNe 
within the dusty, high-density starburst environment and (iii) the extinction 
towards these events. 

Most of the CCSNe in young starbursts like M 82 (Fig. 1; Mattila $\&$ Meikle 
2001, 2002) are expected to be heavily obscured by dust, and therefore remain 
undiscovered by current SN search programmes working at optical wavelengths. 
However, in the near-IR $K_{\rm s}$-band the extinction is greatly reduced and 
the sensitivity of ground-based observations is still good. That is why we are 
using INGRID to carry out an imaging survey of the $\sim$40 most luminous 
nearby ($d<$ 45~Mpc) starburst galaxies visible in the northern hemisphere. 
These galaxies have been selected to have their far-IR luminosities
greater than or 
comparable to those of the two prototypical starbursts, M 82 and NGC 253, 
but excluding galaxies whose far-IR luminosity is powered by a population of 
old stars or an AGN. The expected CCSN rates in these galaxies (Mattila $\&$ 
Meikle 2001) range from around 0.05 yr$^{-1}$ in NGC 253 
($L_{\rm FIR}\sim$10$^{10.3}L_{\odot}$) and 0.2 yr$^{-1}$ in NGC 4038/39 
($L_{\rm FIR}\sim$10$^{10.8}L_{\odot}$) to around 1-2 CCSNe per year in Arp 299 
($L_{\rm FIR}\sim$10$^{11.8}L_{\odot}$). 

Since the average distance to our targets is $\sim$30 Mpc, most of them fit
well within one quadrant of the INGRID detector (2 arcmin $\times$ 2 arcmin).  
This enables us to observe most of the target galaxies using the {\it quadrant 
jitter} method in which the galaxy nucleus is placed in the middle of each 
quadrant of the array in turn. As the three other quadrants contain ``empty'' 
sky, the actual target frames can be used to create a sky frame for each of 
the galaxies and a sky flat field frame for the whole night.  Thus, there is 
generally no need for offset sky images. This improves the observing 
efficiency allowing 2--3 galaxies to be observed  per hour (with 10--20 mins. 
on-source exposure time each). The catch of a clear night is therefore 
20--30 galaxy images in which nuclear CCSNe might be hiding. In Fig. 2 we show 
four individual frames and the final combined frame of Arp 299, a luminous 
infrared galaxy at a distance of 45 Mpc, as an example of a {\it quadrant 
jitter} observation. In Fig. 3 the $K_{\rm s}$ image of another sample
galaxy, NGC 4038/39 (the Antennae, distance 20 Mpc), is shown. In general, to 
increase the number 
of frames and thus the total on-source exposure time per galaxy we perform 
a 4-point dither on each of the quadrants in turn.  Therefore, a full quadrant 
jittering cycle produces typically 16 frames per galaxy, all with different 
offsets. When creating a sky frame the quadrants in which the galaxy is known 
to be are masked out. Thus the number of pixels which are median-combined to 
form a sky frame is 12. The observing procedure is simplified by the use of 
scripts which control the telescope/data acquisition sequence.  The target 
quadrant identification is encoded into the object string in the image header, 
for subsequent use by the data pipeline.

The full sampling of the seeing disk (FWHM $\sim$ 0.7'') by INGRID
(0.24''/pixel) allows us to compare the reduced galaxy images to 
reference frames obtained earlier, 
using image subtraction.  For this we use the Optimal Image 
Subtraction method (Alard $\&$ Lupton 1998, Alard 2000) which derives 
a convolution kernel to match the better seeing image to the 
image with the poorer seeing. It also matches the background differences. 
In Fig. 4 we show an example of image subtraction using a pair of images 
of NGC 253
(distance $\sim$3 Mpc)  observed under different seeing and photometric 
conditions in August 2001 (FWHM = 0.9'') and January 2002 (FWHM = 1.1''). 
Here, 21 different 5.7''$\times$5.7'' regions automatically selected by the 
image subtraction program were used for deriving the convolution kernel. The 
spatial variations of the INGRID PSF were modeled with a 2nd order polynomial 
and the differential background variations with a 1st order polynomial. The 
subtraction residuals are very 
small except for the two bright stars visible in the north-east and 
south-west from the nucleus.  When the image subtraction is carried out with 
a constant kernel solution using just one region centered on the galaxy 
nucleus for deriving the kernel, a slight  PSF variation between the 
frames over the INGRID field of view (4 arcmin $\times$ 4 arcmin) is visible.
Such relative PSF variations can be caused by, e.g., differential 
rotation between the two frames (Alard 2000) as a result of imperfect image 
registration. More image subtraction examples and starburst galaxy images can 
be found at {\it http://astro.ic.ac.uk/nSN.html}. 

The quick, effective reduction and analysis of the SN search data is 
essential for this programme to succeed. Therefore, the INGRID data taken in 
this project is pipeline-processed in near-real time at the telescope
(see Fig. 5).  The IRAF-based pipeline runs 
on one of the Beowulf clusters at the telescope (Greimel et al. 2001). At 
the beginning of the observing run, calibration data is taken to generate a 
bad pixel mask and a 
dome flat field for the image processing. When a new image is taken by the 
data acquisition system it is automatically copied to the cluster and the 
post-pre image subtraction is done. An image grouping task then waits until 
all the exposures for a given galaxy have been taken before feeding the list 
of exposures to the next step. 
Once all the images for one object are on the cluster they are combined to give
an initial sky frame. The images and the initial sky frame are then fed into
a modified de-dithering routine (idedither$\_$qd) from the INGRID quicklook 
package (ingrid$\_$ql) in which the images are registered on a user defined 
star (this will be 
automated in future); a final sky image is created based on quadrant masking; 
sky subtraction, bad pixel masking and flat fielding are done for every 
image; and the processed images are finally combined. The last processing 
step is to apply a World Coordinate System (WCS) solution based on the USNO 
catalogue to the combined 
image. At this point a data analysis task takes over. The new image of the 
object is compared with an archived image from previous runs. The rotation,
shift and scaling are calculated for the image, based on 
user-selected stars. In the future this will be automatically done based on 
the WCS solution. The final step in the data analysis is the image subtraction 
using the Optimal Image subtraction software (ISIS) as described above. The 
subtracted images are then inspected by eye. In addition, the fully reduced 
search images are compared to the existing reference images by blinking.
Apart from automating the various pipeline steps, we are also working on the 
implementation of a web-based interface for the pipeline which will allow
easy steering of the pipeline as well as immediate access to the data
by off-site collaborators. 

Having acquired a complete set of INGRID reference images we estimate a 
probable discovery rate of between 0.4 and 0.8 SNe in each clear night's 
observations (see Mattila $\&$ Meikle 2001, 2002). The newly developed data 
processing pipeline for the on-going nuclear CCSN search on the 
WHT enables an easy 
real-time analysis of the search data.  This is essential for the rapid 
follow-up observations of discovered SNe, in order to determine the nature of 
these events. Near-IR ($JHK$) photometry and spectra will probe both the 
conditions in the immediate circumstellar/interstellar environment of the SN 
and the line-of-sight extinction towards the SN. A large amount of near-IR 
imaging data still needs to be 
collected if we are to detect a sufficient number of obscured SNe to derive a 
statistically significant SN rate in nearby starburst galaxies. Therefore, we 
invite any observers who will be acquiring or have recently acquired K-band 
data of luminous nearby starburst galaxies to take part in the Nuclear SN 
search. Full details of this, and contact information, are given on the 
'Nuclear SN Search' pages at {\it http://astro.ic.ac.uk/nSN.html}. 

\begin{small}
We thank Johan Knapen and Petri V\"ais\"anen for advice and discussions on the observing technique. 
\newline
\end{small}
\vspace{-0.4cm}

References: \\
Alard C., Lupton R.H., 1998, ApJ, 503, 325 \\
Alard C., 2000, A$\&$AS, 144, 363 \\
Chevalier R.A. $\&$ Fransson C., 2001, ApJ, 558, 27 \\
Dahlen T. $\&$ Fransson C., 1999, A$\&$A, 350, 349 \\ 
Fassia A. et al., 2000, MNRAS, 318, 1093 \\
Greimel R., Lewis J.R., Walton N.A., 2001, ING Newsl, 4, 9 \\
Mattila S., Meikle W.P.S., 2001, MNRAS, 324, 325 \\
Mattila S., Meikle W.P.S., 2002, in The central kpc of starbursts and AGN:
the La Palma connection, eds. J.H. Knapen, J.E. Beckman, I. Shlosman and
T.J. Mahoney, A.S.P. Conference Series, 249, 569 \\
Terlevich R., Tenorio-Tagle G., Franco J., Melnick J., 1992, MNRAS, 255, 713 \\
Sullivan M., Ellis R., Nugent P., Smail I., Madau P., 2000, MNRAS, 319, 549 \\
Vink J., de Koter A., Lamers H.J.G.L.M., 2001, A$\&$A, 369, 574 \\

\epsfxsize=8cm
\begin{figure*}
 \vspace{-1cm}
 \hspace{-0.5cm}
 \vbox to60mm{\vfil
 \epsfbox{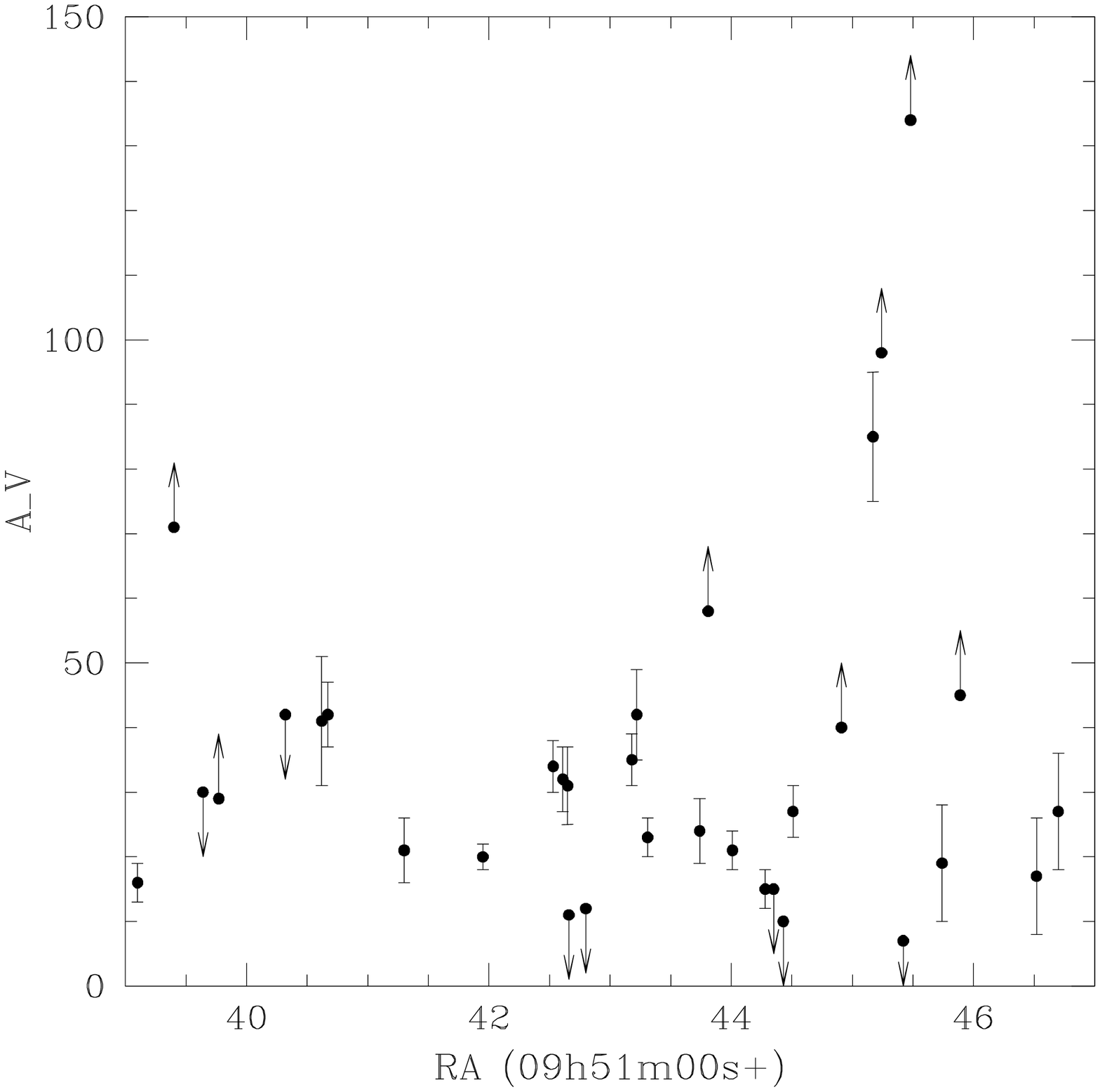}
 \vspace{0cm}
 \hspace{0cm}
\caption{Extinctions ($A_{V}$) towards 33 SNRs in M 82 (from Mattila $\&$
Meikle 2001, 2002).}
 \vspace{1cm}
 \hspace{-0.5cm}
 \epsfxsize=18cm
 \epsfbox{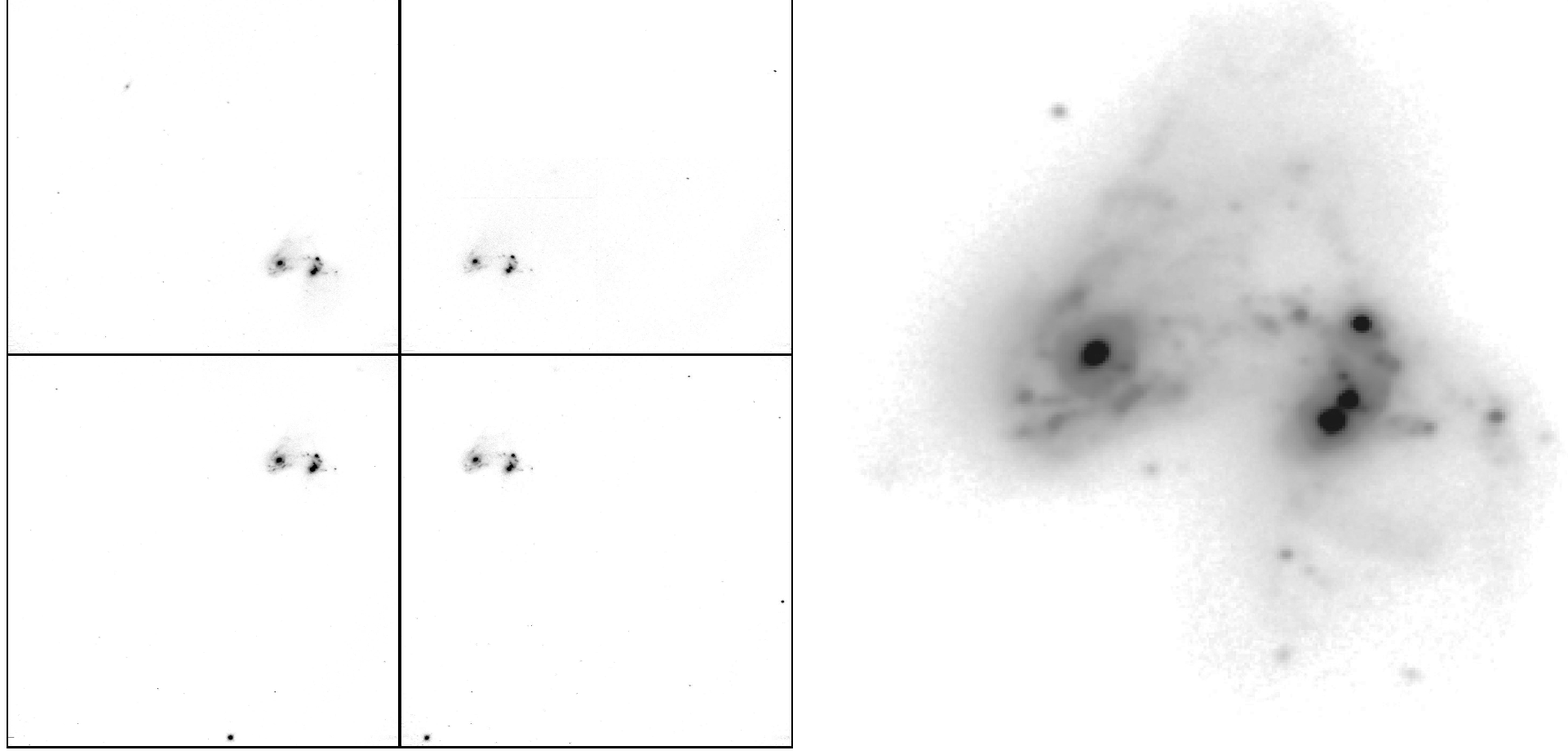}
 \vspace{-16cm}
 \hspace{+2cm}
\caption{$K_{\rm s}$ band images of Apr 299 (North is up and East to the left)
obtained with INGRID at the WHT on January 2nd, 2002. Four ${\it quadrant 
~jitter}$ frames with 10 $\times$ 6 sec. exposure time are shown on the 
left, and the combined frame formed from 32 individual frames (32 mins. total 
exposure time) on the right. The intensity in the combined frame has a square 
root scaling.}
\label{fig2} 
\vfil}
\end{figure*}

\newpage

\epsfxsize=12cm
\begin{figure*}
 \vspace{-12.5cm}
 \hspace{+0cm}
 \vbox to141mm{\vfil
 \epsfbox{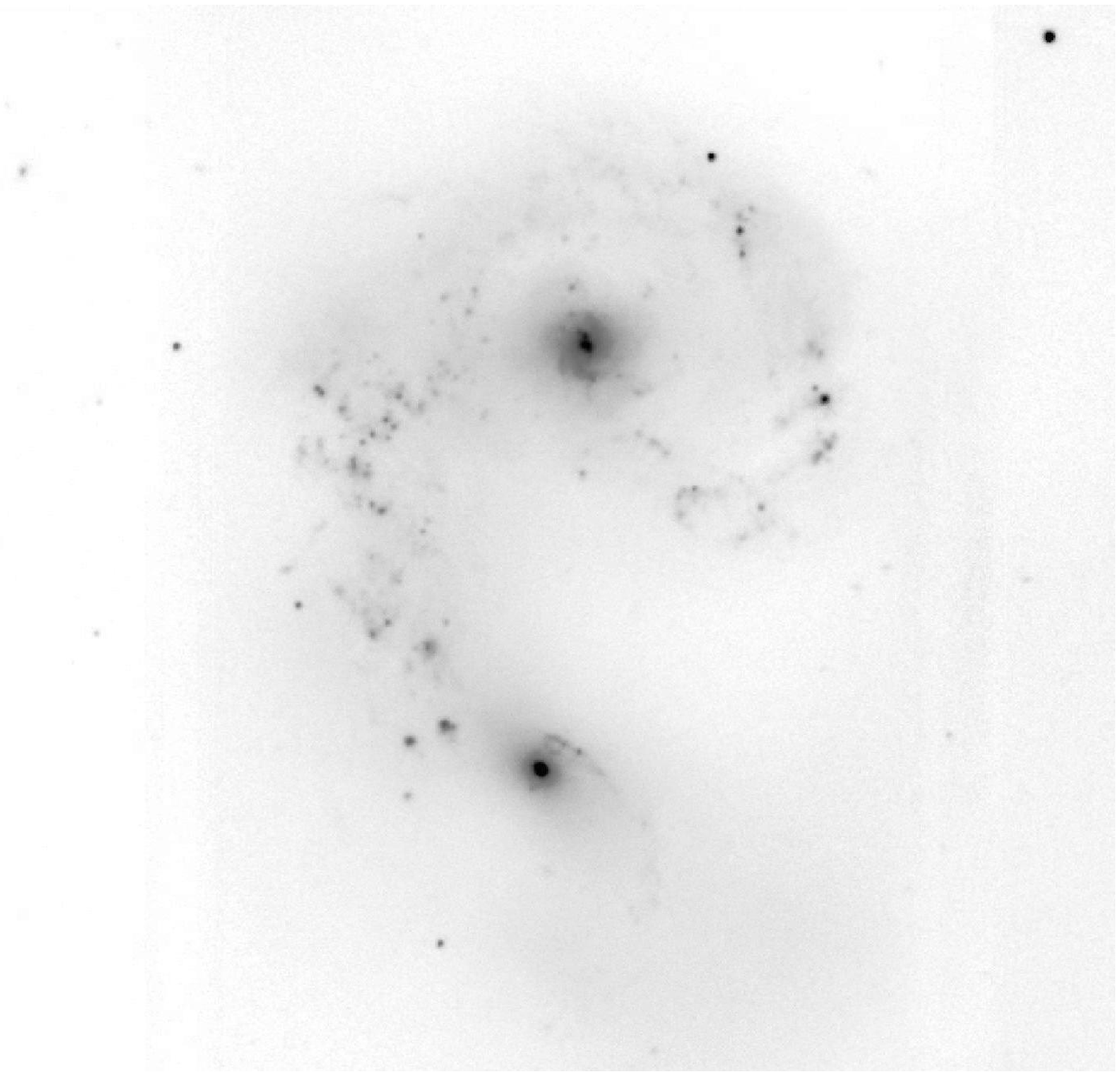}
 \vspace{0cm}
 \hspace{-1cm}
\caption{$K_{\rm s}$ band image of the Antennae, NGC 4038/39 (North is up and 
East to the left) obtained with INGRID at the WHT on January 2nd, 2002
(640 sec. exposure time).}
\epsfxsize=17cm
 \epsfbox{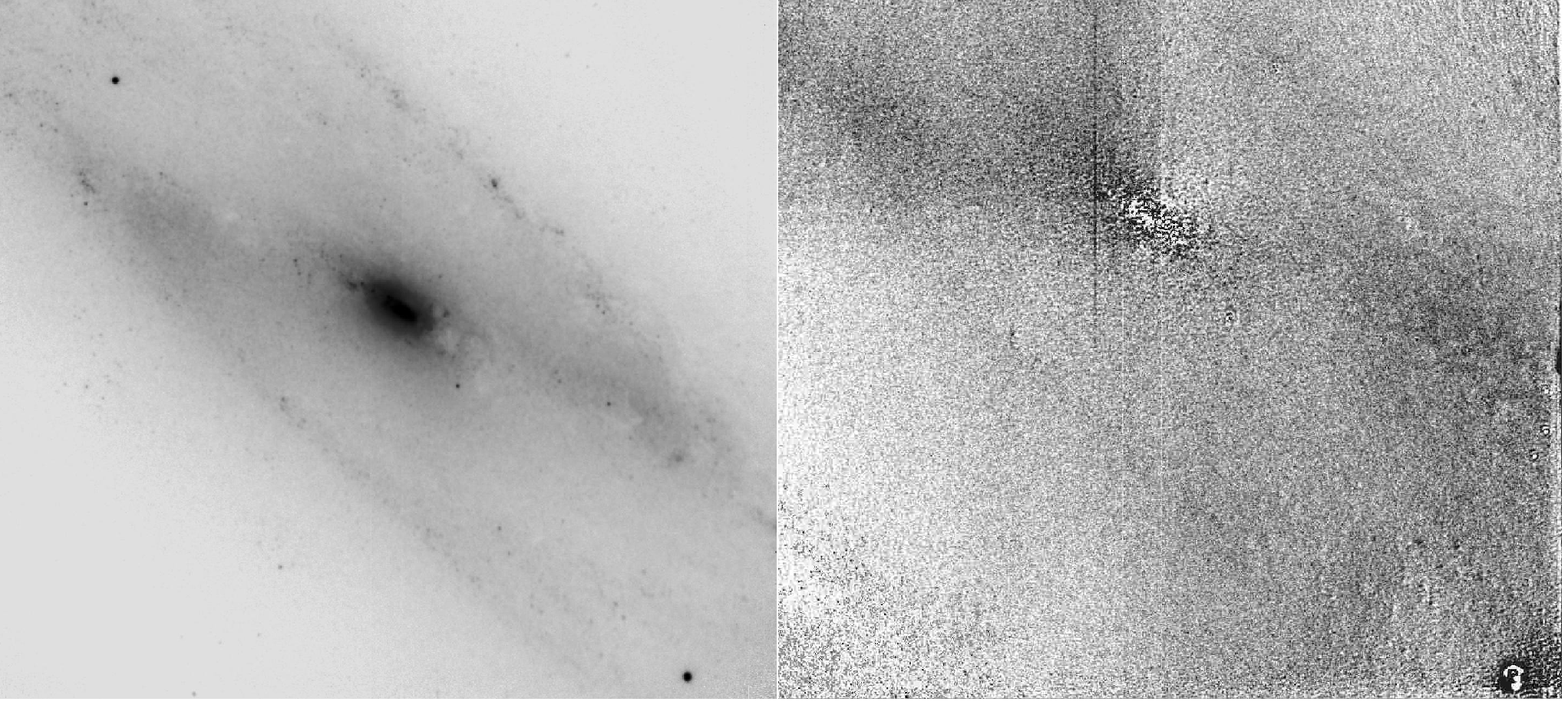}
 \vspace{-15.5cm}
 \hspace{0cm}
\caption{$K_{\rm s}$ band image of NGC 253 (North is up and East to the left)
obtained with INGRID at the WHT on August 31st, 2001 (300 sec. exposure time)
is shown with a square root intensity scaling on the left. The result of 
image matching and subtraction is shown on the right.}
\label{fig4} 
\vfil}
\end{figure*}
\newpage
\epsfxsize=20.5cm
\begin{figure*}
 \vspace{-7cm}
 \hspace{-2cm}
 \vbox to320mm{\vfil
 \epsfbox{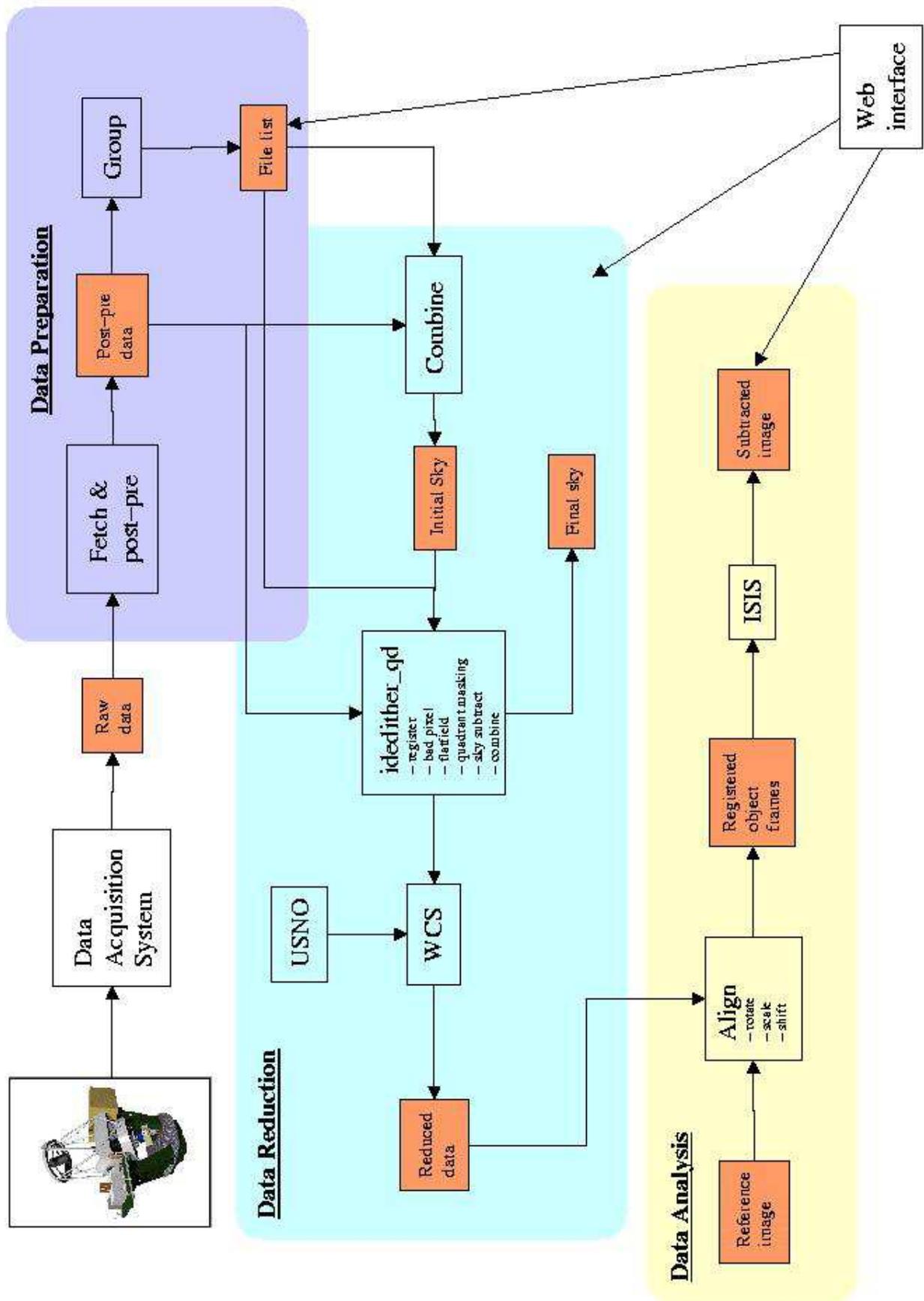}
 \vspace{-2cm}
 \hspace{0cm}
\caption{The real-time data analysis pipeline.}
\label{fig5} 
\vfil}
\end{figure*}
\end{document}